\documentclass[reprint, amsmath, amssymb, aps]{revtex4-1}

\usepackage[pdftex]{graphicx}
\usepackage{wasysym}
\usepackage{amsfonts}
\usepackage{ifsym}
\usepackage{ulem}

\newcommand{\nn}{\nonumber}

\newcommand{\COMMENT}[1]{}

\newcommand{\neqa}{\nonumber\end{eqnarray}}
\newcommand{\la}[1]{\label{#1}}

\newcommand{\<}{{\langle}}
\renewcommand{\>}{{\rangle}}

\newcommand{\re}{\relax{\rm I\kern-.18em R}}

\def\su2{{SU(2)}}

\def\[{\left[}
\def\]{\right]}

\def\({\left(}
\def\){\right)}
\def\[{\left[}
\def\]{\right]}

\def\<{\langle}
\def\>{\rangle}

\def\i2{\frac{i}{2}}

\def\2F1{\,_2{\rm F}_1}

\newcommand{\ft}[2]{{\textstyle\frac{#1}{#2}}}

\usepackage[usenames,dvipsnames]{xcolor}
\usepackage{color}
\usepackage{graphicx}
\usepackage{dcolumn}
\usepackage{bm}
\usepackage{hyperref}

\newcommand{\beq}{\begin{equation}}
\newcommand{\eeq}{\end{equation}}
\newcommand{\beqq}{\begin{equation*}}
\newcommand{\eeqq}{\end{equation*}}
\newcommand\beqa{\begin{eqnarray}}
\newcommand\eeqa{\end{eqnarray}}
\newcommand\beqaa{\begin{eqnarray*}}
\newcommand\eeqaa{\end{eqnarray*}}
\newcommand\bea{\begin{array}}
\newcommand\eea{\end{array}}

\begin{document}

\title{On the collinear limit of scattering amplitudes at strong coupling}

\author{Benjamin Basso$^{\displaystyle\hexagon}$, Amit Sever$^{{\displaystyle\Box}}$ and Pedro Vieira$^{\displaystyle\pentagon}$}

\affiliation{
\vspace{5mm}
$^{\displaystyle\pentagon}$Perimeter Institute for Theoretical Physics,
Waterloo, Ontario N2L 2Y5, Canada\\
$^{\displaystyle\hexagon}$Laboratoire de Physique Th\'eorique, \'Ecole Normale Sup\'erieure, Paris 75005, France\\
$^{\displaystyle\Box}$School of Natural Sciences, Institute for Advanced Study, Princeton, NJ 08540, USA
}

\begin{abstract}

\vspace{3mm}

In this letter we consider the collinear limit of gluon scattering amplitudes in planar $\mathcal{N}=4$ SYM theory at strong coupling. We argue that in this limit scattering amplitudes map into correlators of twist fields in the two dimensional non-linear $O(6)$ sigma model, similar to those appearing in recent studies of entanglement entropy. We provide evidence for this assertion by combining the intuition springing from the string worldsheet picture and the predictions coming from the OPE series. One of the main implications of these considerations is that scattering amplitudes receive equally important contributions at strong coupling from both the minimal string area and its fluctuations in the sphere.

\end{abstract}

\maketitle

\section{Introduction} 
In planar $\mathcal{N}\!=\!4$ Super-Yang-Mills theory, scattering amplitudes and null polygonal Wilson loops are one and the same~\cite{AM,AmplitudeWilson} at any value of the coupling $\lambda = g^2_{YM}N$. Through the prism of the AdS/CFT correspondence, a scattering amplitude can then be viewed as a path integral over the open string configurations that end on a light-like polygon at the boundary of $AdS_5\times S^5$.
At strong coupling, this partition function is dominated by its saddle point which in turn is given by a minimal string area in $AdS_5$.
For the $n$-gluon amplitude this translates into~\cite{AM}
\beq
\mathcal{W}_{n} = e^{-\frac{\sqrt{\lambda}}{2\pi} A_n+\dots} \la{minArea}
\eeq
where $\mathcal{W}_{n}$ is the renormalized amplitude introduced in~\cite{short} and $A_n$ is the corresponding subtracted area of the classical string ending on the $n$-gon. (Both are conformally invariant and {finite} quantities which only depend on the cross-ratios specifying the shape of the boundary null polygon.) Thanks to the integrability of the classical worldsheet theory, the problem of computing this area can be reduced to solving a set of {Thermodynamic Bethe Ansatz} equations with $A_n$ being identified with a free energy of sort, known as the critical Yang-Yang functional~\cite{AGM,AMSV,OPEpaper,short}.
\begin{figure}[t]
\centering
\def\svgwidth{7cm}
\includegraphics[width=0.5\textwidth]{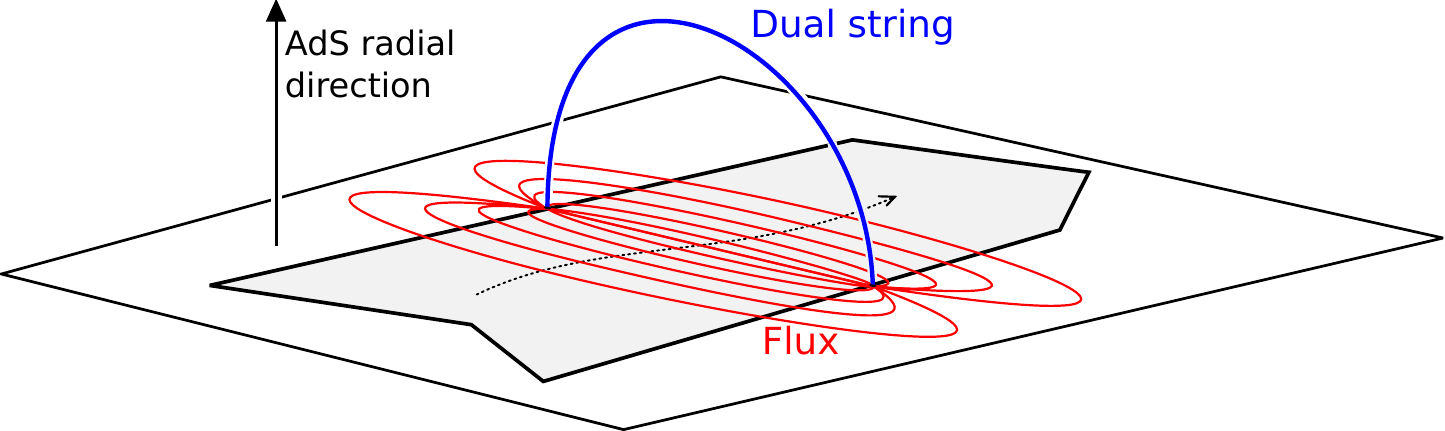}
\caption{A {null} polygon Wilson loop sources a colour flux in the gauge theory whose dual description is that of an open string ending on it. The Wilson loop expectation value can be mapped to a correlator in the flux tube theory or, equivalently, into an open string partition function.  
}\label{OPEpicture}
\end{figure}
Except for that, little is known about scattering amplitudes at strong coupling, that is about the ellipsis in (\ref{minArea}) -- in contrast with the flood of results at weak coupling, see~\cite{review} for a recent review.

{Building upon earlier work \cite{OPEpaper},} we proposed in~\cite{short} an alternative method for computing the open string partition function, at any value of the coupling. In this so-called \textit{pentagon approach} a generic polygon is broken down into a sequence of pentagon transitions $\mathcal{P}$ as \cite{short,data}
\beqa
&&\mathcal{W}_n=\langle 0|  {\mathcal{P}} e^{- {H} \tau_{n-5}+i  {P} \sigma_{n-5} +i  {J} \phi_{n-5}}  {\mathcal{P}} \dots\nn \\
&&\qquad\qquad\qquad\qquad \dots   {\mathcal{P}} e^{- {H} \tau_{1}+i  {P} \sigma_{1} +i  {J} \phi_{1}}   {\mathcal{P}} |0 \rangle \,, \la{OPEoperators}
\eeqa
see~\cite{details} for details. 
This representation is particularly suitable to the analysis of the multi-collinear limit which corresponds to the regime of large $\tau_i$.

Based on this approach as well as on world-sheet considerations, we shall see that at strong coupling the  {\it collinear limit} is governed by the string dynamics in the five sphere. 
More precisely, we will show that in this limit the entire partition function reduces to a correlator of twist operators in the $O(6)$ sigma model, similar to those encountered in the study of entanglement entropy~\cite{CC,CC-AD,Holzhey:1994we}. 

A surprising consequence of this identification and of the strongly coupled dynamics of the $O(6)$ sigma model is an additional exponentially large contribution to $\mathcal{W}_n$ of the same order as the classical area $A_n$.
As we will explain, accommodating for the sphere indeed corrects the minimal area prescription such that
\beq
\log\mathcal{W}_n=-\frac{\sqrt{\lambda}}{2\pi} A_n+ \frac{\sqrt{\lambda}}{48}\frac{(n-4)(n-5)}{n}  +o(\sqrt{\lambda}) \,, \la{new}
\eeq
to leading order at strong coupling. 
More excitingly, the dynamics of the $O(6)$ sigma model also allows us to start unveiling the $\alpha'$ corrections. For the six gluons amplitude, for instance, 
we shall find that 
\beq
\mathcal{W}_6 =  f_6\, \lambda^{-\frac{7}{288}} e^{\frac{\sqrt{\lambda}}{144} - \frac{\sqrt{\lambda}}{2\pi} A_6}(1+O(1/\sqrt{\lambda}))  \la{goal}
\eeq
where the $\lambda$ independent prefactor $f_6$ is a yet to be determined function of the hexagon cross-ratios {$\{\tau, \sigma , \phi\}$}. Computing this function for generic kinematics is beyond the scope of the present paper. However based on the $O(6)$ analysis alone we will predict that in the collinear limit $\tau\gg 1$
\beq
f_6(\tau, \sigma, \phi)\simeq \frac{1.04}{(\sigma^2+\tau^2)^{1/72}} \,, \la{f6}
\eeq
with the critical exponent in this power-law behaviour being related to dimensions of the twist fields mentioned before. 

Finally, we will also see that another face of the strong coupling dynamics of the $O(6)$ sigma model is the breakdown of the string $\alpha'$ expansion for extremely stretched Wilson loops. Namely, we shall observe that for 
exponentially large cross-ratios $\log\tau \gg \sqrt{\lambda} \gg 1$ the open string partition function is fully non-perturbative and governed by the exponentially small dynamical scale of the model. In brief, the emergence of this new scale is the main reason for the richness of the collinear limit at strong coupling.Studying all the various collinear behaviours and their cross-over (as summarized in figure \ref{regions}) is the main subject of this paper.

\section{Pentagons as Twist Operators}

In the collinear limit $\tau_i \gg 1$ the lightest states dominate in (\ref{OPEoperators}).
{At strong coupling, these are the string excitations in the sphere  \cite{AldayMaldacena,Frolov:2002av}, dual to the gauge theory scalars, see e.g. figure 2 in \cite{2pt}. Their dynamics is governed by the $O(6)$ non-linear sigma model and, in particular, }their mass is non-perturbatively generated and exponentially small at strong coupling \cite{AldayMaldacena} 
\beq
m=\frac{2^{1/4}}{\Gamma(5/4)}\,\lambda^{1/8} e^{-\sqrt{\lambda}/4} (1+O(1/\sqrt{\lambda}))\, . \la{massgap}
\eeq
All the other string excitations, i.e., both the AdS and the fermionic modes, have masses of order $1$ at strong coupling~\cite{Frolov:2002av, AldayMaldacena} and {therefore} decouple in the collinear limit.
\begin{figure}[t]
\centering
\def\svgwidth{8.7cm}
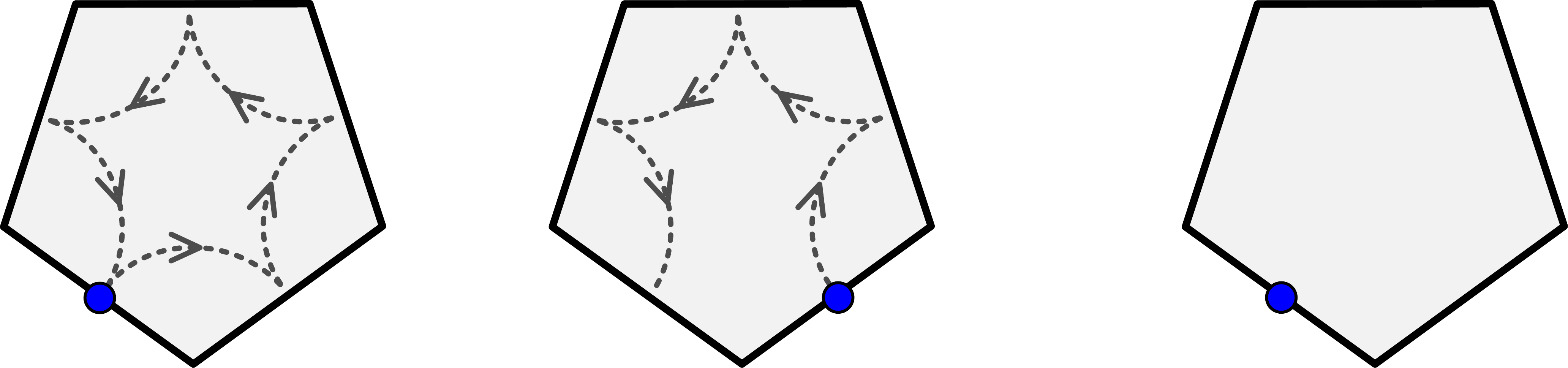
\caption{Under a mirror transformation $\theta\to\theta+i{\pi\over2}$ an excitation is sent from one edge to its neighbour. 
For a pentagon, we need five such transformations to move the excitation all the way around.}\label{mirror}
\end{figure}

This leads us {to} interpret the {strong coupling} collinear limit of (\ref{OPEoperators}) as a correlator in the $O(6)$ model~\cite{DropPhi}
\beq
\mathcal{W}_n\simeq \langle0| \phi_{\pentagon}(w_{n-4})\dots \phi_{\pentagon}(w_1)|0\rangle_\text{O(6)}  \la{O6corr}
\eeq
where {$w_i-w_{i-1}=(\sigma_i,\tau_i)$ and $\phi_{\pentagon}(w)$ are operators} whose matrix elements coincide with the pentagon transitions 
\beq
\<\theta_1',\dots|\phi_{\pentagon}(0)|\theta_1,\dots\>^{j_1,\dots}_{i_1,\dots} = P(\theta_1,\dots|\theta_1',\dots)_{i_1,\dots}^{j_1,\dots} \, . \la{phiP}
\eeq
{Here, $\theta_j$ are the} usual hyperbolic rapidities {parametrizing the scalars' relativistic dispersion relation} while the indices refer to the $O(6)$ polarizations of the incoming and outgoing multi-scalar states. 

\begin{figure}[t]
\centering
\def\svgwidth{8.5cm}
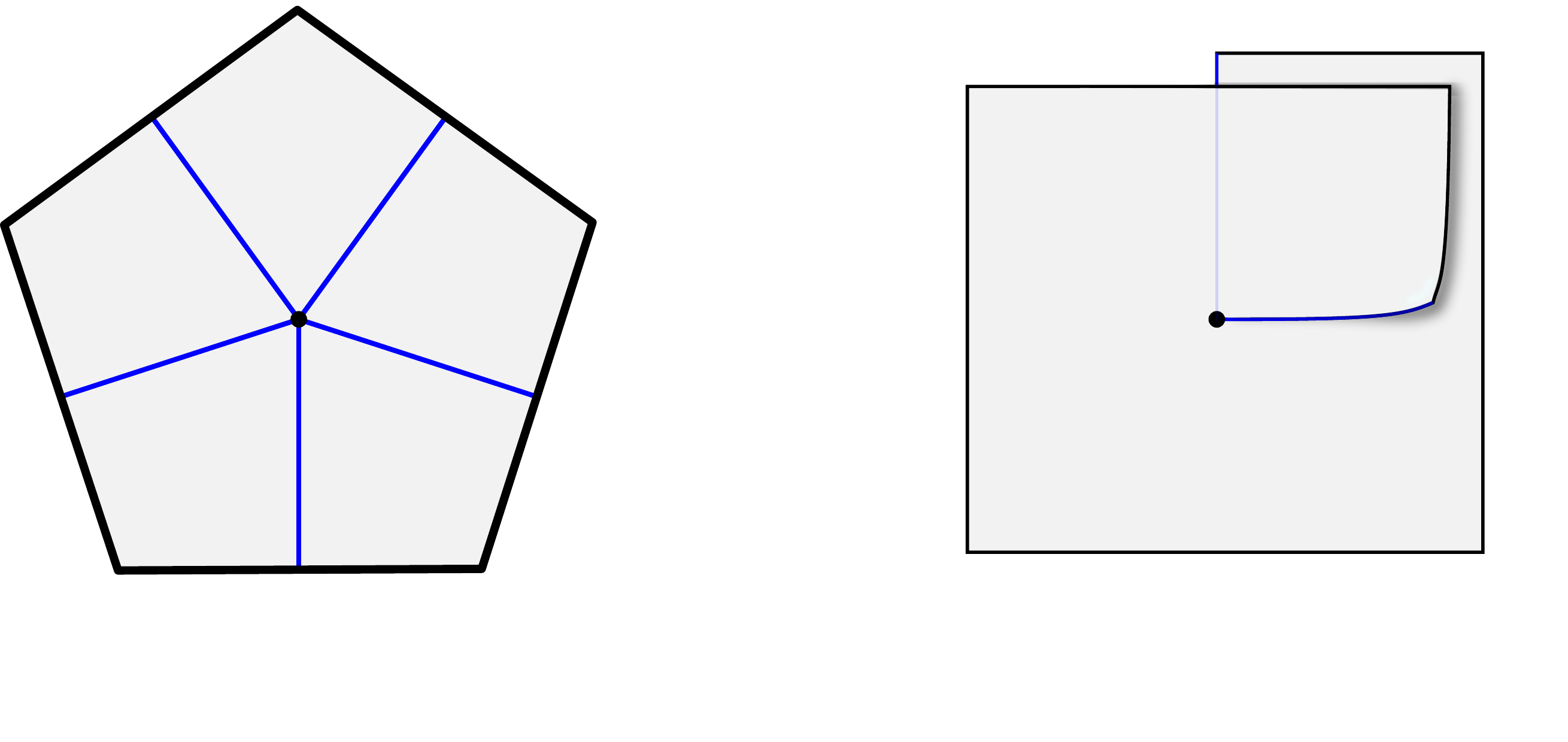
\caption{(a) The world-sheet of the string ending on a pentagon can be viewed as made out of five quadrants. 
(b) Equivalently, we can engineer these five quadrants starting from the square by inserting the twist operator $\phi_{\pentagon}$. This one generates an excess angle of $2\pi/4$.
}\label{Pentagonhas5regions} 
\end{figure}

The clue about what the operator $\phi_{\pentagon}(w)$ is comes from the observation that one needs to perform $5$ so-called mirror rotations (equivalently Euclidean boost) $\theta \to \theta+5 i \pi/2$ to go around the pentagon \cite{short}, see figure \ref{mirror}. 
This should be contrasted with the more standard monodromy for conventional local operators which involves $4$ such transformations only. This hints that 
the effect of the operator $\phi_{\pentagon}(w)$ is to generate a conical  excess angle $\frac{1}{4}\times 2\pi$ around $w$. Such fields are not entirely new and belong to a broad class of operators known in the CFT literature as {\it twist operators} \cite{Dixon:1986qv}. 
Most directly relevant for {our discussion} is their appearance in the context of entanglement entropy \cite{CC,CC-AD}.
There, such operators were introduced to study QFTs on $k$-sheeted Riemann surfaces with branch points being viewed as twist operators with excess angle $\varphi = 2\pi(k-1)$ in the replica theory.
Our case is somewhat special in that it requires a ``fractional number of sheets" since $k=5/4$ for a pentagon, see figure \ref{Pentagonhas5regions}. {As} further evidence that this identification is correct, 
one can verify that the pentagon transitions in the right-hand side of (\ref{phiP}) satisfy the axioms for the form factors of twist operators as spelled in~\cite{CC-AD} with $k=5/4$.

The above picture can also be understood more directly from the worldsheet analysis. From our previous discussion it follows that the partition function (\ref{OPEoperators}) receives{, in the collinear limit, its} dominant contribution from the sphere. This means that we can write (up to normalization)
\beqa
\mathcal{W}_n \simeq \int \mathcal{D}X  e^{-\delta S_{NG}} \,\delta(X^2-1)
\eeqa
where
\beqa
\delta S_{NG}= \frac{\sqrt{\lambda}}{4\pi} \int d^2z\, \sqrt{g} g^{\alpha\beta}\partial_\alpha X \cdot \partial_\beta X 
\eeqa
is the expansion of the Nambu-Goto action to quadratic order in the sphere embedding coordinates $X$ and $g_{\alpha\beta}$ is the induced metric of the classical minimal surface in $AdS$~\cite{Induced}. We thus face the problem of computing the partition function of the $O(6)$ sigma model on the minimal surface. From the low-energy viewpoint, this surface looks everywhere flat, except for a few points where the curvature is concentrated. Indeed the induced metric in the collinear limit is approximately
\beq
ds^2\ {\simeq}\ [P(z)\bar P(\bar z)]^{1/4} dz d\bar z
\eeq
where $P(z)=\prod_{j=1}^{n-4}(z-z_j)$ is the auxiliary polynomial entering the Pohlmeyer description of the minimal surface~\cite{AGM}. {In agreement with the pentagon picture, we see that there are} $n-4$ marked points around which we have a conical excess of $2\pi\times \ft{1}{4}$. Following~\cite{CC}, the partition function in this geometry can be recast as a correlator of $n-4$ twist operators as~(\ref{O6corr}). 

\section{OPE as Form Factor Expansion} 

\begin{figure}[t]
\centering
\def\svgwidth{8cm}
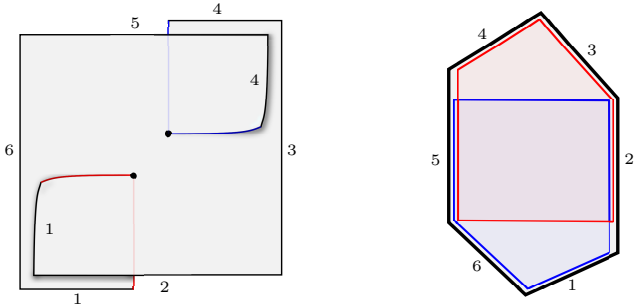
\caption{At strong coupling, the hexagon Wilson loop in the collinear limit is given by a correlator of two twist operators in the $O(6)$ sigma model (on the left), corresponding to the two pentagons in its decomposition ({on the right}).}\label{Hexagonhas6regions}
\end{figure}

As elaborated above, at strong coupling, the collinear limit is governed by the dynamics of the $O(6)$ sigma model whose physics is strongly coupled. As such, at the moment, the only available tool for studying this regime in a controllable way is the pentagon approach \cite{short}. In this section we will focus on the simplest possible case, the hexagon $\mathcal{W} = \mathcal{W}_6$, see figure \ref{Hexagonhas6regions}. 

Given the relativistic invariance of the O(6) sigma model, the Wilson loop can only depend (in the collinear limit) on 
  the dimensionless Lorentz invariant distance
\beq
{z} \equiv m \sqrt{\sigma^2+\tau^2} \,. \la{xdef}
\eeq
For any value of $z$, the correlator in~(\ref{O6corr}) can then be written as
\beq
\mathcal{W}= \!\!\! {\sum _{n\text{ even}} \!  \int\! \frac{\prod_i\! d\theta_i}{n!\,(2\pi)^{n}}} |P(0|\theta_1,\!...,\theta_{n})_{i_1,...,i_{n}} |^2\, e^{-z\! \sum\limits_{k=1}^{n}\! \cosh\theta_k }\la{sum} \,.
\eeq
This is the familiar form factor expansion, which simply follows from inserting the resolution of the identity between consecutive operators in~(\ref{O6corr}). Alternatively, from the Wilson loop point of view, this sum stands for the truncation of the full OPE series to the scalar subsector in the strong coupling limit~\cite{footnote3}. 
(We refer the reader to the conclusions for a discussion of the corrections to~(\ref{sum}).)

\begin{figure}[t]
\centering
\def\svgwidth{3.5cm}
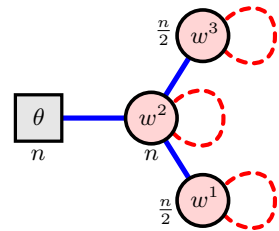
\caption{The pattern of auxiliary rapidities arising in the construction of the matrix part has a group theoretical interpretation. The three sets of rapidities can be identified with the three nodes of the $O(6)$ dynkin diagram. The occupation numbers are fixed such that the overall state with $n$ particles belongs to the $O(6)$ singlet representation. The cartoon depicted here is related to~(\ref{mpart}) by identifying the solid lines with $f$'s and the dashed lines with $g$'s. 
}\label{auxiliaryroots}
\end{figure}

As illustrated in \cite{2pt}, the pentagon transitions can be factored out into a dynamical factor and a so-called matrix part taking care of the matrix structure of these objects. {Namely} 
\beq
 |P(0|\theta_1,\dots,\theta_{n})_{i_1,\dots,i_{n}} |^2 = \Pi_\text{dyn} {\times} \Pi_\text{mat}
\eeq
Working out these contributions (most notably the matrix part) in a systematic fashion is a fascinating problem which we will report elsewhere \cite{toAppear}. The main conjecture arising from this analysis is that $\Pi_{\text{mat}}$ is a rational function of the (differences of) rapidities $\theta_j$ which admits a very simple integral representation involving $2n$ auxiliary rapidities. Namely,
\beqa
\Pi_\text{mat}=\frac{1}{n!(\frac{n}{2}!)^2}\int\limits_{-\infty}^{+\infty}  \frac{dw^{1}_1 \dots dw^{1}_{\frac{n}{2}} dw^{2}_1 \dots dw^{2}_{n} dw^{3}_{1} \dots dw^{3}_{\frac{n}{2}}}{(2\pi )^{2n}} \nn \\ 
 \times\frac{\prod\limits_{i< j} g(w_i^{1}-w_j^{1})\prod\limits_{i< j} g(w_i^{2}-w_j^{2})\prod\limits_{i< j} g(w_i^{3}-w_j^{3})}{\prod\limits_{i,j} f(\frac{2}{\pi}\theta_i-w_j^{2})\prod\limits_{i,j} f(w_i^{1}-w_j^{2})\prod\limits_{i,j} f(w_i^{3}-w_j^{2})}\qquad   \la{mpart}
\eeqa
with $f(x)=x^2+1/4$ {and} $g(x)=x^2(x^2+1)$. A self-explanatory depiction of the matrix part integral is shown in figure \ref{auxiliaryroots}.
We should stress that for any fixed number of particles{, $n$,} the integrals over the auxiliary roots can be {straightforwardly} evaluated by residues. In particular, for $n=2$, one easily verifies {in} this way that
\beq
\Pi_\text{mat}^{n=2} = \frac{3\pi^4}{2(\theta_{12}^2+\pi^2)(4\theta_{12}^2+\pi^2)}\, ,  \qquad \theta_{12} = \theta_{1}-\theta_2\, .
\eeq
Finally, the dynamical part takes the factorized form
\beq 
\Pi_\text{dyn} \!=\! \mu^n \prod\limits_{i<j} F(\theta_i-\theta_j) \,, \qquad \mu=\frac{2\Gamma(\frac{3}{4})}{\sqrt{\pi} \Gamma(\frac{1}{4})} \,,
\eeq 
with 
\beq
 F(\theta)=\frac{8\, \theta  \tanh \left(\frac{\theta }{2}\right) \Gamma \left(\frac{3}{4}+\frac{i \theta }{2 \pi }\right) \Gamma \left(\frac{3}{4}-\frac{i \theta }{2 \pi
   }\right)}{\pi\,  \Gamma \left(\frac{1}{4}+\frac{i \theta }{2 \pi }\right) \Gamma \left(\frac{1}{4}-\frac{i \theta }{2 \pi }\right)} \,.
\eeq

In concluding this section, it is important to stress that the result (\ref{sum}) is a novel exact result for scattering amplitudes. It holds at strong coupling and in the collinear limit with the Lorentz invariant distance (\ref{xdef}) held fixed, but otherwise arbitrary. It is inherently non-perturbative.

\section{Long and short distance analysis} \la{LS}

Two very interesting regimes one might want to analyze in greater detail are the IR regime where $ {z}\gg 1$ and the UV regime where $ {z}\ll 1$. The former is straightforwardly extracted from (\ref{sum}) since it is dominated by the vacuum contribution
\beq
\mathcal{W}=1 + O(e^{-2 {z}})\, . \la{fullOPE}
\eeq
The first deviation is controlled by the 2-particle integral which was previously analyzed in \cite{2pt}. The trivialization of the Wilson loop in this limit is in perfect agreement with the expected behaviour of scattering amplitudes in the collinear limit. We note that it is achieved for $\tau$ much greater than the Compton wavelength $1/m$ of the lightest excitations. ({In other words, we only reach (\ref{fullOPE}) for highly stretched Wilson loops whose cross-ratios take extreme values $\log\tau \gg \sqrt{\lambda}$.})

As usual with the form factor expansion, it is much more challenging to analyze the UV regime $ {z}\ll 1$. The point is that the higher-particle terms in the sum~(\ref{sum}) are no longer suppressed at small $z$. Instead, they typically explode and the full series~(\ref{sum}) must be resummed.
The two-particle contribution, for instance, displays the logarithmic behaviour
\beq
\mathcal{W}_{2\text{-pt}}  {\xrightarrow[z\to0]{} r \log(1/z)+s \log\log(1/z) }+ t + o(1)\, ,
\eeq
with $r  \simeq 0.031, s \simeq -0.055$ and $t \simeq -0.008$.
The expectation -- which we confirmed numerically on few examples -- is that the $n$-particle contribution should follow the same trend and diverge as $\log({1/z})^{n/2}$ at small $z$. Clearly, without further information, it is challenging to predict what the true $z$ dependence will be upon re-summing all contributions in~(\ref{sum}). Fortunately, the twist-field interpretation introduced before sheds light on this issue and provides us with a physical picture for what the result should be, as we now explain. 

\begin{figure}[t]
\centering
\includegraphics[scale=0.8]{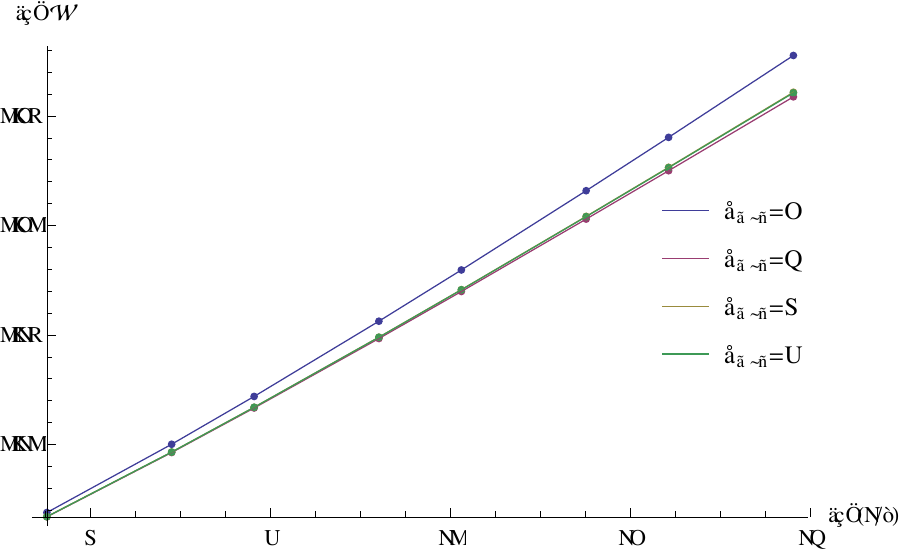}
\caption{Plot of $\log\mathcal{W}$ truncated to $n_{\textrm{max}}$ particles for $z$'s as small as $10^{-6}$ and as large as $1/250$. The 2-particle approximation corresponding to $n_{\textrm{max}}=2$ (i.e., the upper line) already yields a reasonable estimate of the exact result; this is not unusual for such form factor representations,
see e.g.~\cite{CC-AD}. For the range of $z$'s under consideration it is clear that summing the contributions up to $6$ particles is plenty; indeed the curves with $n_{\textrm{max}}=6$ and $n_{\textrm{max}}=8$ (i.e., the middle lines) are already indistinguishable in the figure. The numerics corroborates the power law behaviour described in the text.}\label{numerics}
\end{figure}

The hexagonal Wilson loop is computed by a correlator of two twist operators in the $O(6)$ sigma model. In the short distance limit, these two operators are fused according to their OPE. Given that each operator has the effect of producing a conical excess of $\pi/4$, a pair of close by pentagons should act as an effective `hexagon' operator producing a conical excess of $\pi/2$.
In other words, we expect the short distance OPE to be given by
\beqa
\phi_{\pentagon}(\sigma,\tau)\phi_{\pentagon}(0,0) \sim \frac{\log(1/z)^B}{z^{\ \!\!A}} \,\phi_{\hexagon} (0,0) \,, \la{OPEbehavior}
\eeqa
where $A=2\Delta_{\pentagon}-\Delta_{\hexagon}=2\Delta_{5/4}-\Delta_{3/2}$ with $\Delta_k$ the dimension of the twist field (with excess angle $2\pi \times (k-1)$). The latter dimension has been known for a long time~\cite{Knizhnik:1987xp} and reads
\beq\Delta_k =\frac{c}{12}\(k-\frac{1}{k}\) \, , \la{DeltaEq}
\eeq where $c$ is the central charge. In our case $c=5$ since the short distance CFT is that of $5$ free massless (Goldstone) bosons. This leads to the sharp prediction $A=1/36$ for the leading power law behaviour. 

The critical exponent $B$ might look less familiar at first sight, as it is absent from the OPE of primaries in standard CFTs. It controls however a celebrated  logarithmic enhancement which comes about because we are dealing with an asymptotically free theory and because our operators receive anomalous dimensions. (This is very well known from QCD and $B = -(2\gamma_{\pentagon}-\gamma_{\hexagon})/(2\beta_0)$ when expressed in terms of one-loop anomalous dimension and beta function coefficients, see e.g.~\cite{Weinberg2}.) Unfortunately,  to our knowledge, these anomalous dimensions are not yet available from direct QFT computations. Still, it is possible to argue for a possible relation between them and the free energy of the $O(6)$ sigma model. We defer the details of the argument to the appendix and quote here the main conjecture $B=-3A/2$. 

All in all, once inserted into the correlator (\ref{O6corr}) the OPE (\ref{OPEbehavior}) generates the short distance behaviour
\beq
\mathcal{W}(z) = \frac{C}{z^{1/36}\log(1/z)^{1/24}}+\dots \la{prediction}
\eeq
where $C$ is a constant which reflects the freedom in adopting different normalizations for the twist fields. For the problem at hand, the physical normalization is set by the collinear limit. Namely, it is unambiguously fixed by the long distance asymptotics~(\ref{fullOPE}) which is equivalent to $\left<0\right|\phi_{\pentagon}(w)\left|0\right> = 1$ by clustering.
Because this condition is imposed in the IR, where the non-perturbative physics dominates, it is challenging, if not impossible, to fix $C$ from the CFT directly.  
\begin{figure}[t]
\centering
\includegraphics[scale=0.6]{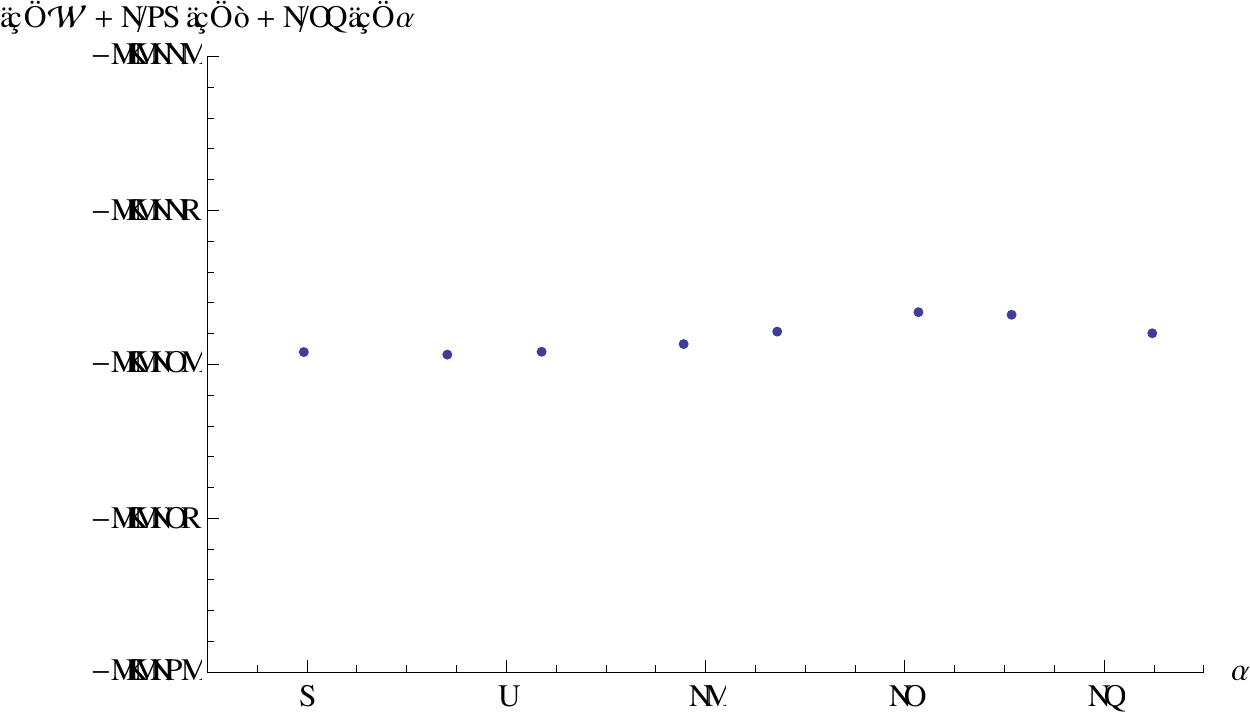}
\caption{Plot of $\log \mathcal{W}$ as a function of the two-loop running coupling $\alpha$. The latter is related to $z$ through the RG equation $\alpha-1/4 \log \alpha= \log(1/z)$. Our estimate for $\log\mathcal{W}$ is based on the numerical evaluation of the series representation (\ref{sum}) (truncated to eight particles, see figure \ref{numerics}). We see that it approaches a constant at small $z$ after subtracting the leading and sub-leading logarithmic behaviours. This leads to the prediction (\ref{Cvalue}) for the constant $C$. 
}\label{numerics2}
\end{figure}
What we can do, however, is to fix our constant $C$ numerically, through the exact series representation (\ref{sum}) truncated at some large number of particles. Dealing with the multi-dimensional integrals in (\ref{sum}) is numerically challenging. One way to do it is by Monte-Carlo, along the lines of \cite{Yurov:1990kv} which analysed a similar (yet simpler) form factor sum related to a correlator in the 2d Ising model. In figure \ref{numerics} we represent the numerical evaluation of the OPE series for increasingly small values of $z$. As depicted in figure \ref{numerics2}, we observe that, once we subtracted the leading and subleading logarithmic behavior, $\log \mathcal{W}$ does approach a constant value (which we can identify with $\log C$). In this way we read
\beq
\log C\simeq -0.01  \la{Cvalue}
\eeq
for the constant. It would be interesting to improve the numerics and get $C$ with higher precision. Even better, it would be great if we could compute it analytically from the OPE sum (\ref{sum}).

\section{Cross-over and Classical Enhancement}

We are now in position to explain the prediction~(\ref{goal}), (\ref{f6}) for the $\alpha'$ expansion of the six-gluon amplitude. Essentially what we want to show is that the short-distance $O(6)$ result (\ref{prediction}) is enough to fix the prefactor dressing the minimal area prediction~(\ref{minArea}) in the collinear limit. It is well known that in this limit the classical area $A_{6}$ falls off exponentially fast with $\tau$~\cite{OPEpaper}
\beq
A_{6} = O(e^{-\sqrt{2}\tau})\, , \qquad \tau \gg 1\, ,
\eeq
and similarly for the $n$-gluon area $A_n$ in the multi-collinear limit $\tau_{i}\gg 1$.
This behaviour is {most clearly} understood {by} recalling that the $AdS_5$ modes, which control the physics of the minimal surface, \textit{all} have masses of order $O(1)$. {(The lightest ones have} mass $\sqrt{2}$~\cite{Frolov:2002av,AldayMaldacena}, see e.g. figure 2 in \cite{2pt}). Therefore, whatever survives in the collinear limit is necessarily captured by the prefactor dressing the minimal area prediction~(\ref{minArea}).

\begin{figure}[t]
\centering
\def\svgwidth{9cm}
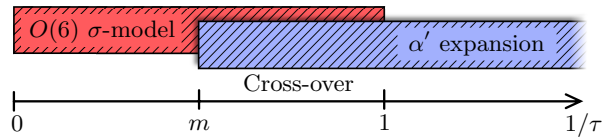
\caption{Cross-over between the non-pertubative regime captured by the $O(6)$ {$\sigma$-}model and the perturbative region covered by the string $\alpha'$ expansion. The two relevant scales in the problem are the non-perturbative mass gap $m \ll 1$ and the ``UV" scale $\sim 1$ at which the massive modes of the string enter. The former sets the boundary of the $\alpha'$ expansion while the latter is an upper bound for the validity of the low-energy description. The two descriptions should agree over the range $1 \ll \tau \ll 1/m$ which corresponds to the short distance regime of the $O(6)$ model on the one hand and to the collinear limit of the $\alpha'$ expansion on the other. 
}\label{regions}
\end{figure}

That the aforementioned prefactor is non-trivial in this limit directly follows from our previous analysis. The main point is that regardless of how big $\tau$ is, from the string $\alpha'$ expansion point-of-view, we always end up in the short-distance regime $z\ll 1$ of the $O(6)$  model. Indeed, for fixed $\tau$ and very large $\lambda$, the dimensionless distance $z$ given by (\ref{xdef}) is very small. In other words, $z\ll 1$ is the cross over domain between the non-perturbative regime $z\sim 1$ analyzed in this paper and the perturbative regime of the string worldsheet theory, as illustrated in figure~\ref{regions}.

This being said, it is straightforward to convert the short-distance result~(\ref{prediction}) into the prediction~(\ref{goal}), (\ref{f6}).  It literally amounts to matching the latter against the former using the expressions (\ref{xdef}) and (\ref{massgap}) for the distance $z$ and the mass gap $m$. (In more technical terms this is the usual conversion between RG improved and conventional perturbative expansions).

What is perhaps the most surprising outcome of all this analysis is the semi-classical enhancement stemming from the dynamics in the sphere. Namely, we see that the contribution from the sphere {is visible} already at the leading order in the $\sqrt{\lambda}$ expansion. {Technically,} this is a consequence of the fact that the twist fields carry scaling dimensions.
{Namely}, our correlators are all dimensionless by construction and thence all distances come multiplied by $m$. In the short distance limit the overall dependence on the mass of the correlators can then be directly read off the OPE of the twist fields. In the case of $n$-gluon scattering, we would have $n-4$ pentagons that fuse together into an object with excess angle $\varphi = 2\pi \times \frac{n-4}{4}$. Keeping track of the mass dependence only we would then write
\beq
\underbrace{\phi_{\pentagon}\dots \phi_{\pentagon}}_ {n-4} \sim m^{-(n-4)\Delta(\frac{5}{4})+\Delta(\frac{n}{4})}\phi_{\varphi}\, .
\eeq
This immediately yields
\beq
\log{\mathcal{W}_{n}} \sim \frac{(n-4)(n-5)}{12 n}\log{(1/m)} + AdS\, \,  \textrm{part}\, , \la{WnEnd}
\eeq
or equivalently~(\ref{new}).

As a final remark, let us add that the O(6) model can also be used to predict the pre-factor dressing the strong coupling result~(\ref{WnEnd}) in the multi-collinear limit ($\tau_i\gg1$) for any $n$-gon. 
To leading order at strong coupling, it should relate to the correlation function $\<\phi_{\pentagon}(w_1)\ldots \phi_{\pentagon}(w_{n-4})\phi_{\varphi}(\infty)\>_\text{CFT}$ in the free theory which depends non-trivially on the ratios of distances between the points $w_i=(\tau_i,\sigma_i)$. Following~\cite{Lunin:2000yv}, its computation should lead to a beautiful mathematical problem in classical Liouville theory which would be fascinating to analyze.

\section{Conclusions}

In this paper we start unveiling the structure of scattering amplitudes at strong coupling in planar $\mathcal{N}=4$ SYM theory beyond the minimal area paradigm. 
We learned that scattering amplitudes are schematically of the form
\beq
\mathcal{W}=C \,\lambda^B\, e^{A\sqrt{\lambda}} \, .
\eeq
The leading term, $A$, receives both a contribution from $AdS_5$ and from $S^5$, with the former admitting a classical description as opposed to the latter which is fully non-perturbative.

The subleading term, $B$, is a constant that only depends on the number of gluons. It comes solely from the sphere. This type of $\lambda$ prefactor was not unexpected; similar pre-factors were found before for other Wilson loops using localization. The most notable example is the circular Wilson loop \cite{kostya}. There, the exponent $B$ was related to a simple counting of zero modes. Our $B$ is not so different (although a bit more complicated) in the sense that it is uniquely determined by the low energy degrees of freedom.

Finally we have the prefactor $C=C^{(0)}+C^{(1)}/{\sqrt{\lambda}}+\dots $ which depends non-trivially on the geometry. It receives all kind of contributions and it is a fascinating problem to understand them thoroughly. In this paper we proposed that the collinear limit provides a good starting point for its study. We have seen that the leading behaviour of $C$ in this limit is fully captured by the $O(6)$ sigma model. 
We could now envisage completing this story by progressively taking into account all different corrections away from the collinear limit. These are essentially of two kinds. One amounts 
for correcting the integrals over the scalars by taking into account $1/\sqrt{\lambda}$ corrections to the pentagon transitions and dispersion relation. From the world-sheet {description}, such corrections can be interpreted as irrelevant deformations of the low energy effective theory (i.e. of the $O(6)$ sigma model). These type of corrections will typically lead to power-law suppressed contributions in $\tau$. Being suppressed by $1/\sqrt{\lambda}$, they contribute to $C^{(1)}$ only. The other kind of corrections are related to the string massive modes and are exponentially suppressed at large~$\tau$. These are important ones as they will contribute to $C^{(0)}$ already. In the OPE set-up, they come from including all the excitations into our sums. This should amount, in the worldsheet theory, to computing the full one-loop determinant around the classical solution, which is a daunting but fascinating problem.

In the end, one might hope that this prefactor takes a particularly inspiring form from the integrability point of view, akin to the critical Yang-Yang functional governing the minimal area. If so, one could imagine bootstrapping it completely from the knowledge of the first few corrections away from the collinear limit, mimicking somehow the successful bootstrap program at weak coupling~\cite{Lance}. 

To conclude, in this letter we have seen how strong coupling dynamics might challenge our intuition about scattering amplitudes, or their dual description in terms of Wilson loops, already in such a seemingly simple regime as the collinear limit. The rich behaviour we observed directly reflected the strong IR effects on the dual world-sheet which come about because the colour flux tube of the theory is infinite and its spectrum effectively gapless at strong coupling. These features will survive beyond the planar limit and are common to some other strongly coupled flux tubes, see e.g.~\cite{Aharony:2009gg}.
\\
\\
{\it Acknowledgements:} We thank L.~Dixon, V.~Kazakov, J.~Toledo, E.~Yuan and especially J.~Maldacena for enlightening discussions and suggestions. Research at the Perimeter Institute is supported in part by the Government of Canada through NSERC and by the Province of Ontario through MRI. A.S was supported in part by U.S. Department of Energy grant DE- SC0009988.

\section{Appendix}
In this appendix we provide evidence for the conjecture $B=-3A/2$ presented in section~\ref{LS}.

To fix $B$ we need the dimension of the twist operator. This in turn is equivalent to computing energy on a cylinder (by the operator-state correspondence). Before acting with the twist operator, the ground state of the system is the vacuum of the replica theory. After the conformal map to the cylinder, this looks like $k$ independent copies of a cylinder of length $L$ and energy $E_{\textrm{vac}}(L)$. The effect of the twist operator is to join these copies together such as to form a single cylinder of length $k L$. Denoting by $\Delta_k$ the dimension of the operator and by $E_k(L)$ the energy of the corresponding state, this translates into
\beq
\Delta_k=\frac{L}{2\pi} \( E_k(L)-k E_\text{vac}(L) \) \, . \la{Dapp}
\eeq
Since in the case at hand we are interested in the vacuum energy, we have that $E_k(L) = E_\text{vac}(kL)$.

Equation~(\ref{Dapp}) is easily seen to reproduce the dimension~(\ref{DeltaEq}) of the twist operator, see~\cite{Lunin:2000yv} and below, when we sit at the UV fixed point. More importantly for us, we also expect it to hold true if we weakly perturb the system and start flowing off the conformal point. Assuming this is case, we can read the one-loop dimension of the twist operator from the subleading correction to the vacuum energy at small $L$. In an asymptotically free theory, the latter is well-known to admit the expansion 
\beq
E_\text{vac}(L)=- \frac{\pi}{6L}\(c+ \frac{\delta c}{2\beta_0\log(mL)}+\dots\) \, , \la{EvacL}
\eeq
with $c$ the UV central charge and $\delta c$ a coefficient governing the one-loop correction (note that in stringy notation ${1}/{\log(mL)}\propto 1/\sqrt{\lambda}$). Clearly the former reproduces (\ref{DeltaEq}) while the latter gives us the one-loop anomalous dimension coefficient
\beq
\gamma_k = \frac{\delta c}{12} \(k-\frac{1}{k}\)\, .
\eeq
Hence computing $B = -(2\gamma_{5/4}-\gamma_{3/2})/(2\beta_0)$ boils down to determining $\delta c/(2\beta_0)$. In principle, it is straightforward to obtain the energy~(\ref{EvacL}) using the thermodynamic Bethe ansatz (TBA) equations for the vacuum energy. These are known for the $O(N)$ sigma models at any $N$~\cite{Fendley:1999gb}. In practice, however, solving the TBA at small $L$ is difficult. An alternative approach is to use the large $N$ analysis carried out in~\cite{Balog}. Using this result it is possible to argue~\cite{AndreiBen} that $\delta c= 3\beta_0 c$ for any $N$. (As further evidence, we checked this relation numerically against the TBA numerics~\cite{Balog2} for the $O(3)$ and $O(4)$ sigma model.) Given that $A$ and $B$ differ by replacing $c$ by $-\delta c/(2\beta_0)$, the conjecture $B=-3A/2$ immediately follows.


\begin{thebibliography}{99}
\bibitem{details}

The main objects in (\ref{OPEoperators}) are the pentagon operators ${\mathcal{P}}$ which act on the flux-tube Hilbert space. The coordinates $\tau_i,\sigma_i,\phi_i$ parametrize the cross-ratios of the polygon and $H,P,J$ are the flux-tube Hamiltonian, momentum and angular momentum operators, see \cite{short,data} for more details.
\bibitem{DropPhi}
The scalars have no angular momentum and {thus} the angle variables $\phi_j$ drop from (\ref{OPEoperators}).
\bibitem{footnote3} 
{
Here we use the relativistic normalization which differs from the one in \cite{short,2pt} by a simple rescaling of the measure and pentagon transitions, see section 8 of \cite{2pt}. 
%\color{red} More precisely, with respect to the conventions of \cite{short,data,2pt} we have $\mu_\text{here}=1=1/g\times \mu_\text{there}/\mu$ where $\mu$ is the same constant appearing in (\ref{Pdyn}) and $g=\sqrt{\lambda}/4\pi$ as usual. Similarly, $P_\text{here}=\dots$. 
}
\end{thebibliography}
\end{document}